\documentclass[preprint,onecolumn,amsmath,showkeys]{revtex4}

\usepackage{graphicx}
\usepackage{dcolumn}
\usepackage{bm}
\usepackage[T1]{fontenc}

\begin{document}

\title{Landau free energy of ferroelectric crystals by thermodynamic integration}

\author{Gr\'egory Geneste$^1$\footnote{electronic address : gregory.geneste@ecp.fr}}
\address{$^1$ Laboratoire Structures, Propri\'et\'es et Mod\'elisation des Solides, CNRS-UMR 8580,
Ecole Centrale Paris, Grande Voie des Vignes, 92295 Ch\^atenay-Malabry Cedex, France\\}

\begin{abstract}
Using Molecular Dynamics simulations based on the effective hamiltonian developed by Zhong, Vanderbilt and Rabe [Phys. Rev. Lett. {\bf 73}, 1861 (1994)] (and fitted on first-principles calculations only), the technique of the thermodynamic integration is applied to barium titanate. It allows to compute the difference of free energy between macroscopic states with different polarizations, from the thermal averages of the forces acting on the local modes. This is achieved by performing molecular dynamics under the constraint of fixed polarization. The Landau free energy is thus interpreted as a potential of mean force. The thermodynamic integration directly gives access (numerically) to the Landau free energy of barium titanate as a function of $\vec P$, without any assumption on its analytical form. This technique, mainly used in computational chemistry, allows to make a direct connection between phenomenological theories and atomic-scale simulations. It is applied and validated to the case of BaTiO$_3$ under fixed volume. The results are compared to available phenomenological potentials of the litterature.
\end{abstract}


\maketitle

\section{Introduction}
Ferroelectric (FE) solids have been the subject of many theoretical studies for a long time. In particular, the nature of their phase transitions, displacive versus order-disorder, and the microscopic mechanisms that might be responsible for them, the space and time correlation of dipoles, the damping of the soft mode in the high-temperature phases, have attracted very soon much interest since these phenomena are of fundamental interest to understand the deep nature of ferroelectricity.

From a phenomenological point of view, FE phase transitions are well described in the framework of the Landau theory. This theory assumes the existence of a so-called Landau free energy $\tilde{F}(N,V,T,\vec P)$, that can be seen as an "incomplete" thermodynamic potential\cite{iniguez2001}:

\begin{equation}
\tilde{F}(N,V,T,\vec P) = - k_B T ln \tilde{Z}(N,V,T,\vec P),
\end{equation}

in which $\tilde{Z}(\vec P)$ is an incomplete partition function defined by 
\begin{equation}
 \tilde{Z}(N,V,T,\vec P) = \sum_{i/\vec P} e^{- \epsilon_i/k_B T}
\end{equation}
In this expression, the summation is extended over the microscopic states $(i)$ with energy $\epsilon_i$ exhibiting a polarization equal to $\vec P$. In Landau theory, the equilibrium order parameter $\vec P$ is obtained as the value minimizing the Landau free energy.

The Landau theory has been successfully applied to BaTiO$_3$ (BTO) by Devonshire in the fiftees\cite{devonshire49,devonshire51,devonshire54}. He expanded the Landau free energy in power series up to sixth order in the three components of the polarization $P_x$, $P_y$ and $P_z$. Assuming a linear dependence with temperature of the coefficient of the $P^2$ term (as stated by Landau theory), he showed that the three phase transitions of barium titanate can be described with a relatively simple expression of the free energy. Since this pionnering work, Landau theory has been applied to barium titanate in many other forms than bulk (films\cite{shirokov2007}, wires\cite{hong2008}) and to many other ferroelectrics\cite{ferrobook}.

The direct calculation of the Landau free energy requires two things:

(1) performing simulations with fixed polarization, which can be achieved easily within constrained Molecular Dynamics (with modified equations of motion obtained from lagrangian formalism), or by applying a suitable external electric field\cite{iniguez2001}.

(2) calculating the thermodynamic potential.

This last point is the most difficult since standard simulation techniques such as Monte Carlo (MC) or Molecular Dynamics (MD) do not give a direct access to the free energy, that can not be attained as the thermal average of some microscopic quantity. These two techniques only give access to macroscopic states of the system that are {\it minima} of the thermodynamic potential.

The coefficients of the free energy expansion for barium titanate have been calculated from the first-principles derived effective hamiltonian of Zhong {\it et al}\cite{zhong94,zhong95} in an elegant way by I$\tilde{n}$iguez {\it et al}\cite{iniguez2001} through Monte Carlo simulations under applied external electric field. Although Monte Carlo only gives access to the minima of the free energy, these authors used the external field to displace the minima in the 3-D space (P$_x$, P$_y$, P$_z$) and get access to the coefficients of the expansion and their temperature dependence. In fact, their approach could have been extended to obtain the derivatives of the free energy with respect to the polarization at various points, and thus, by integration, get access to the free energy as a function of $\vec P$ without any assumption on its analytical form. The approach we propose in this article is formally equivalent to this point.

Indeed, some efficient techniques do exist to get directly insight into the thermodynamic potential (or more precisely into differences of thermodynamic potential), such as umbrella sampling\cite{umbrella1,umbrella2}, free energy perturbation\cite{zwanzig1954} or thermodynamic integration\cite{kirkwood}. The basic idea of thermodynamic integration is to find a "reaction coordinate" $\lambda$ joining continuously two macroscopic states $\lambda_0$ and $\lambda_1$, and calculate the derivative of the free energy with respect to $\lambda$ (normally accessible within constrained MD) along this path. This derivative can thus be integrated to obtain the difference of free energy between $\lambda_0$ and $\lambda_1$:

\begin{equation}
\tilde{F}(\lambda_1) - \tilde{F}(\lambda_0) = \int_{\lambda_0}^{\lambda_1} \left\{ \frac{\partial \tilde{F}}{\partial \lambda}(\lambda)\right\}_{NVT} d \lambda,
\end{equation}

or even the whole free energy profile along the path. In this formula, $\tilde{F}(\lambda)$ is the incomplete thermodynamic potential defined above.

In most cases, this yields:

\begin{equation}
\tilde{F}(\lambda_1) - \tilde{F}(\lambda_0) = \int_{\lambda_0}^{\lambda_1} \left\langle \frac{\partial U}{\partial \lambda}\right\rangle_{NVT} d \lambda,
\end{equation}

in which $U$ is the potential energy\cite{darve2001}.

This technique is widely used in computational chemistry and chemical physics, two fields in which scientists are interested in calculating free energy profiles along a reaction coordinate $\lambda$ to obtain the activation free energy of a chemical reaction or its equilibrium constant (related to the standart change in free energy). The free energy for a given value of $\lambda$ (referenced to that with $\lambda_0$) is interpreted as the "potential of mean force" on the reaction coordinate

\begin{equation}
\tilde{F}(\lambda) - \tilde{F}(\lambda_0) = - \int_{\lambda_0}^{\lambda} 
\left\langle f(\lambda') \right\rangle d \lambda',
\end{equation}

{\it i.e.} the potential related to the averaged force acting on the reaction coordinate $\lambda$ (usually $- f(\lambda)$ is the external force that must be applied to maintain the reaction coordinate invariant).

In our case, if we choose for $\lambda$ the polarization and take as starting point ($\lambda$=0) the paraelectric state of BTO at some given temperature, it is thus formally possible to get access to the free energy for any value of $\vec P$, {\it i.e.} the Landau free energy of the ferroelectric crystal. Thus we interpret the Landau free energy as a potential of mean force, calculated as a function of the polarization, considered as the reaction coordinate.
Chemists have been using this technique for a long time to calculate reaction free energies and activation free energies along paths defined by a reaction coordinate\cite{ti1}. In this paper, we transpose the thermodynamic integration technique to the field of ferroelectricity.

\section{Theoretical background}

\subsection{Effective Hamiltonian}
We adopt the formalism of the Effective Hamiltonian\cite{zhong94,zhong95}. In this approximation, the number of degrees of freedom of the perovskite (15 per ABO$_3$ unit cell) is reduced to 6 per ABO$_3$ unit cell, chosen to be local collective displacements constructed from the lowest-energy eigenvectors of the force constant matrix. Two kinds of such modes are considered: the "local modes" $\vec u_i$, that roughly represent the dipole in each unit cell, and the "displacement modes" $\vec v_i$, describing the inhomogeneous strain. The use of such degrees of freedom has revealed to be efficient and sufficient to describe correctly the thermodynamics of BTO (in particular its complex sequence of phase transitions).
In terms of these relevant degrees of freedom, the potential energy is the Effective Hamiltonian $H^{eff}(\left\{\vec u_i\right\},\left\{\vec v_i\right\},\left\{\eta_l^H\right\})$, $\eta_l^H$ being the homogeneous strain tensor.

The local modes $\vec u_i$ are related to the polarization by:

\begin{equation}
\vec P = \frac{Z^{*}}{\Omega} \sum_i \vec u_i,
\end{equation}
where $\Omega$ is the volume of the supercell and $Z^{*}$ the effective charge of the local modes.

We define an average local mode $\vec u$:
\begin{equation}
\vec u =  \frac{1}{N} \sum_i \vec u_i,
\end{equation}
in which $N$ is the number of unit cells in the supercell.
Fixing $\vec P$ is thus equivalent to fix $\vec u$ and we use $\vec u$ for convenience in the following (we calculate the free energy as a function as $\vec u$). $\vec u$ has the dimension of a displacement.

\subsection{Molecular dynamics under fixed polarization}
Our approach in this work is based on molecular dynamics (MD). The first-principles derived hamiltonian of Zhong {\it et al} can be solved either via Monte Carlo simulations or Molecular Dynamics, that should yield similar results under the ergodicity hypothesis. We have already validated the MD method with the effective hamiltonian on barium titanate\cite{md} and showed that its complex sequence of phase transitions as well as the temperature evolution of the polarization and strain can be reproduced successfully as it is in Monte Carlo. This is achieved by combining the Nos\'e-Hoover\cite{nose84,nose86,hoover85} (to fix the temperature) and the Parinello-Rahman\cite{pr} (to fix the pressure/stress) algorithms. More details about this molecular dynamics are given in Appendix B.

Anyway, to calculate differences of free energy along a path indexed by polarization (reaction coordinate), it is necessary to perform MD under fixed polarization. Additional forces have to be added in the dynamical equations of motion that ensure the constraint $\sum_i \vec u_i = N \vec u$. These additional forces can be found in a simple manner starting from the lagrangian formalism. Let us call $L$ the lagrangian of the system without constraint, and $L'$ that of the system with constraint.
\begin{equation}
L' = L - \sum_{\alpha} \xi_{\alpha} (\sum_i u_{i\alpha} - N u_{\alpha})
\end{equation}
in which $\xi_{\alpha}$ are Lagrange multipliers and $\alpha=x,y$ and $z$.
Applying the Lagrange equations on $L'$ gives a new set of equations of motion for the local modes:
\begin{equation}
m_{lm}\frac{d^2 u_{i \alpha}}{dt^2} =  f^{lm}_{i \alpha} -  \xi_{\alpha},
\end{equation}
in which $m_{lm}$ is the mass associated to the local modes\cite{md} and $f^{lm}_{i \alpha}$ the $\alpha$-component of the force acting on the $i^{th}$ local mode.
Summing over $i$ gives immediately the value of the Lagrange multiplier along $\alpha$:

\begin{equation}
N \xi_{\alpha}(t) = \sum_i f^{lm}_{i \alpha}(t)
\end{equation}
Maintaining the polarization fixed is thus equivalent to add to the force acting on each local mode (and at each time step) a term equal to (minus) the (space-)average of the forces. The {\it total} external force that must be applied to maintain $\vec P$ fixed is thus $- \sum_i \vec f^{lm}_i$.

$\sum_i \vec f^{lm}_i$ will be thus qualified in the following of this paper as the (internal) force "acting on the polarization" (one must apply its opposite to the system so as to maintain the polarization invariant).

The equations of motion of the displacement modes $\vec v_i$ are not affected by the constraint.

\subsection{Thermodynamic integration}
The application of the technique of the thermodynamic integration yields in our case:
\begin{equation}
\label{int_thermo}
\tilde{F}(\vec u) - \tilde{F}_0 = - \oint \sum_{i} \left\langle \vec f^{lm}_i \right\rangle (\vec u') . d\vec u',
\end{equation}

in which a continuous path is assumed from $\vec u = \vec 0$ to $\vec u$.
$\tilde{F}$ is the free energy of the entire supercell, and the summation is over all the local modes of the supercell. N,V and T are assumed constant along this path. A proof of this equation is given in Appendix A. For each point of the path, the thermal average of the forces is calculated under fixed $\vec u$.
In the previous formula, $\tilde{F}_0 = \tilde{F}(\vec u = \vec 0)$.

In the particular case of a linear path along the $\alpha$-axis:
\begin{equation}
\tilde{F}(u_{\alpha}) - \tilde{F}_0 = - \int_{0}^{u_{\alpha}} 
 \sum_{i} \left\langle f^{lm}_{i \alpha} \right\rangle  (u'_{\alpha}) d u'_{\alpha},
\end{equation}
in which the calculation is performed along the $\alpha$ direction ($u_{\alpha}$ is the $\alpha$ component of $\vec u$, and $u_{\beta \neq \alpha}$=0 along the path).

The formula can be written in a local form:
\begin{equation}
\vec \nabla_{\vec u} \tilde{F} (\vec u) = - \sum_{i} \left\langle \vec f^{lm}_i \right\rangle (\vec u)
\end{equation}

The Landau free energy is thus the potential of mean force acting on the polarization. It is the integral of minus the thermal average of the total force that is acting on the polarization ({\it i.e.} the total external force that must be added to maintain the constraint).
Written as an infinitesimal variation
\begin{equation}
dF (\vec u) = -  \sum_i \left\langle \vec f^{lm}_i \right\rangle (\vec u) . d \vec u,
\end{equation}
the right hand-side member can be interpreted as a (thermodynamic) work and the link with the basic laws of thermodynamics is easy (see Appendix C).

If we define a free energy per unit cell $\Phi (\vec u)$, and an average force per unit cell also $\left\langle \vec f \right\rangle$, we get the very simple expression:
\begin{equation}
\left\langle \vec f \right\rangle = - \vec \nabla_{\vec u} \Phi
\end{equation}

We have thus to perform constrained MD simulations at different values of $u_{\alpha}$, and calculate in each case the thermal average of the sum of the forces acting on the local modes. This is why MD is well suited to Thermodynamic Integration since the forces are directly available.

\section{Computational details}
We perform Molecular Dynamics (MD) based on the Effective Hamiltonian of Ref.~\onlinecite{zhong95} using the code described in Ref.~\onlinecite{md}. We use a 12$\times$12$\times$12 supercell with periodic boundary conditions for the thermodynamic integration. For the direct MD simulations, we also use this size of supercell, excepting below the phase transition, in which finite-size effects are responsible for rapid reversal of the polarization: in that case we use 14$\times$14$\times$14 and 16$\times$16$\times$16 supercells (10 K below the phase transition) to improve the quality of the numerical results.
The equations of motion are solved within the well-known Verlet algorithm. We use a Nos\'e-Hoover\cite{nose84,nose86,hoover85} thermostat to control the temperature. The paths over which the integration is performed are sampled with a step of $\Delta u =$ 0.001a$_0$ (a$_0$ is the BTO lattice constant obtained from the LDA\footnote{a$_0$ = 7.46 a.u. from Ref.~\onlinecite{zhong95}}). Each simulation with fixed polarization is performed on 10$^5$ steps, the thermal averages being computed on the 50000 last steps. The time step is fixed to 2 $\times$ 10$^{-15}$ s.
All the free energy curves presented herafter are performed at {\it fixed strain tensor}, and each curve corresponds to a fixed temperature, to ensure we calculate a (Helmholtz) free energy curve $\tilde{F}(N,V,T,\vec u)$.

The thermodynamic integration along the $\alpha$ direction is simply performed, from zero to $u_{\alpha}=n \Delta u$, via: 
\begin{equation}
\label{rectangle}
\nonumber
 \int_{0}^{n \Delta u} 
 \sum_{i} \left\langle f^{lm}_{i \alpha} \right\rangle  (u'_{\alpha}) d u'_{\alpha} =
 \Delta u \sum_{k=0}^{n} \sum_{i} \left\langle f^{lm}_{i \alpha} \right\rangle (k.\Delta u),
\end{equation}
the thermal average being performed for each value of k under fixed polarization ($= k \Delta u$).


\section{Thermodynamics of BaTiO$_3$ under fixed volume.}
We apply our method to the case of BaTiO$_3$ under fixed volume. This system has been studied very soon, and it was suggested from Landau theory\cite{devonshire51,drougard55} that BTO under fixed volume might undergo a second-order phase transition at the Curie temperature. Several experiments\cite{drougard55,stern61} have confirmed that clamped BaTiO$_3$ has the coefficient of $P^4$ (in the Landau-Devonshire expansion) positive, making the paraelectric-ferroelectric transition of second order.

\subsection{Isotropic strain}
We first focus on BTO constrained to have a cubic unit cell, with a strain tensor frozen to the values found in its high-temperature cubic phase from direct MD simulations\cite{md}: $\eta_1=\eta_2=\eta_3=$0.0121 and $\eta_4=\eta_5=\eta_6$=0.0. This corresponds to a lattice constant of 3.995 {\AA}. We note that these strains result from the application of a negative pressure -4.8 GPa ({\it i.e.} positive stress tensor $\sigma_{xx} = \sigma_{yy}=\sigma_{zz}=$4.8 GPa), which is the one used by Zhong {\it et al}\cite{zhong95} to correct the underestimation of the lattice constant by the Local Density Approximation (LDA). Moreover, the strain is defined here with respect to this cubic LDA ground state (lattice constant 7.46 a.u.).

\begin{figure}[htbp]
    {\par\centering
    {\scalebox{0.30}{\includegraphics{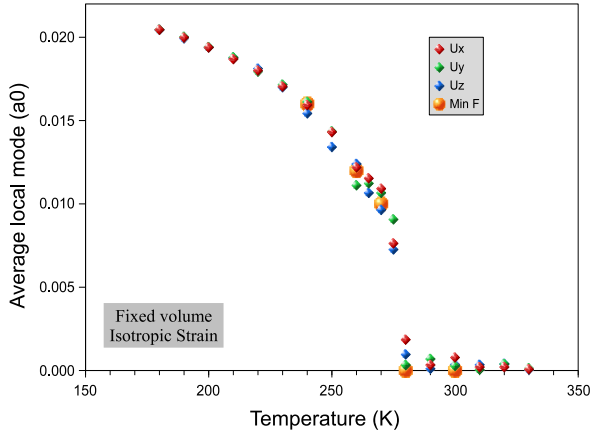}}}
    \par}
     \caption{{\small Temperature evolution of the three components of the polarization from direct MD simulations (without constraint). The polarization minimizing the Landau free energy is shown in orange circles (5 calculations surrounding the phase transition). A 12$\times$12$\times$12 supercell is used except below the transition where the supercell size is 14$\times$14$\times$14 and 16$\times$16$\times$16 between T$_c$-10K and T$_c$.}}
    \label{cubique2}
\end{figure}

Using the Nos\'e-Hoover MD method, we first perform simulations to determine the temperature evolution of the polarization and the possible phase transitions. In fact we only find one phase transition in that case, from cubic (high T) to rhombohedral (low T) (Fig.~\ref{cubique2}) at $\approx$ 280 K. Zhong {\it et al}\cite{zhong95} have already noticed this point through MC simulations on BTO, but with a homogeneous strain tensor frozen entirely to zero. Moreover, the transition seems to be second order or at least weakly first-order (the fluctuations just below the Curie temperature are very large and finite-size effects make the polarization flip from a direction to another, which makes it difficult to discriminate).

\begin{figure}[htbp]
    {\par\centering
    {\scalebox{0.30}{\includegraphics{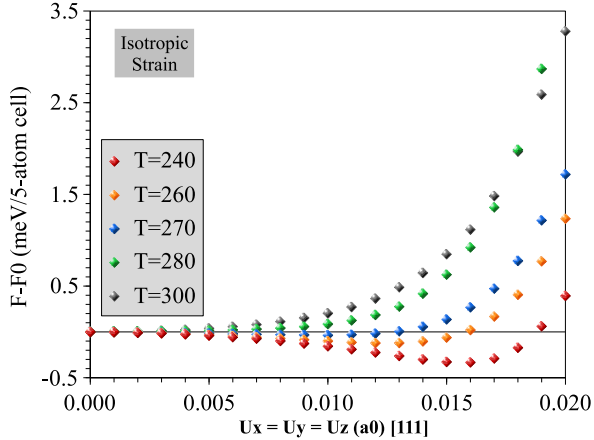}}}
    \par}
     \caption{{\small Landau free energy of cubic BTO (fixed volume) along [111] for 5 temperatures. F0 is the free energy of the paraelectric state for each temperature. The five curves plot $\tilde{F}(N,V,T,\vec u) - \tilde{F}(N,V,T,\vec u=\vec 0)$ as a function of $u_x$ ($=u_y=u_z$) along [111].}}
    \label{cubique1}
\end{figure}

We now turn to thermodynamic integration for a set of selected temperatures surrounding the Curie temperature: T=320, 300, 280, 260 and 240 K. We perform the integration along two paths starting from $\vec u = \vec 0$ and going to $0.025 a_0 \vec e_x$ (along [100]) and to $0.025 a_0 (\vec e_x+\vec e_y+\vec e_z)$ (along [111]). The free energy, resulting from the integration of the thermal averages of the total forces along the second path ({\it i.e.} along [111]), is shown on Fig.~\ref{cubique1}. The minima of the free energy are reported on Fig.~\ref{cubique2} (orange circles) for the five temperatures: 
{\it as expected, the results obtained by direct Molecular Dynamics correspond to the minima of the free energy.}
We have also computed the profile of the free energy along the [100] direction, leading to globally higher free energies than along [111].

Interestingly, the rhombohedral ferroelectric phase obtained at low temperature does not exhibit shear strain (since shear strain is fixed to zero) as it would be the case for a stress-free crystal. Instead, we observe a shear stress, as can be seen on Fig.~\ref{stress} (inset): the non-diagonal components of the stress tensor are not zero at low temperature and decrease to zero at the transition.

\subsection{Tetragonal strain}
\begin{figure}[htbp]
    {\par\centering
    {\scalebox{0.30}{\includegraphics{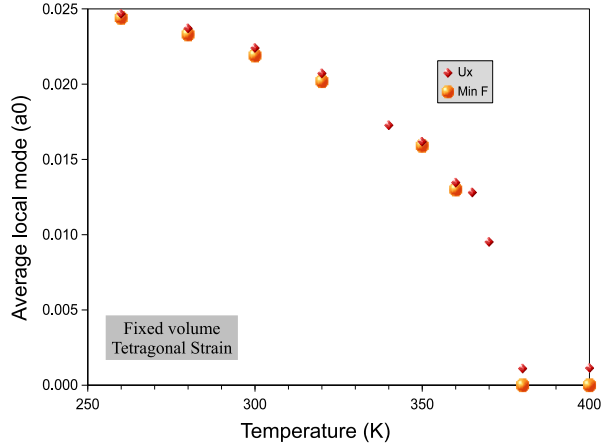}}}
    \par}
     \caption{{\small Temperature evolution of the $x$ component of the polarization from direct MD simulations (without constraint) in the case of tetragonal BTO (tetragonality along $x$). The $y$ and $z$ components are not shown since they are zero. The polarization minimizing the Landau free energy is shown in orange circles (8 calculations). A 12$\times$12$\times$12 supercell is used except below the transition where the supercell size is 14$\times$14$\times$14 and 16$\times$16$\times$16 between T$_c$-10K and T$_c$.}}
    \label{tetra2}
\end{figure}

As another example, we freeze the strain to the values found in the tetragonal phase at T=280 K, {\it i.e.}
$\eta_1=0.01755,~\eta_2=\eta_3=0.01027$ and $\eta_4=\eta_5=\eta_6=0.0$ (as in the previous case, these strains corresponds to a negative hydrostatic pressure of -4.8 GPa at this temperature). The corresponding lattice constants are a=4.017 {\AA} and b=c=3.988 {\AA}. We first perform direct molecular dynamics simulations from T=260 to 400K (Fig.~\ref{tetra2}). The system slowly evolves from ferroelectric to paraelectric, the transition temperature being at $\approx$ T=375 K.

\begin{figure}[htbp]
    {\par\centering
    {\scalebox{0.30}{\includegraphics{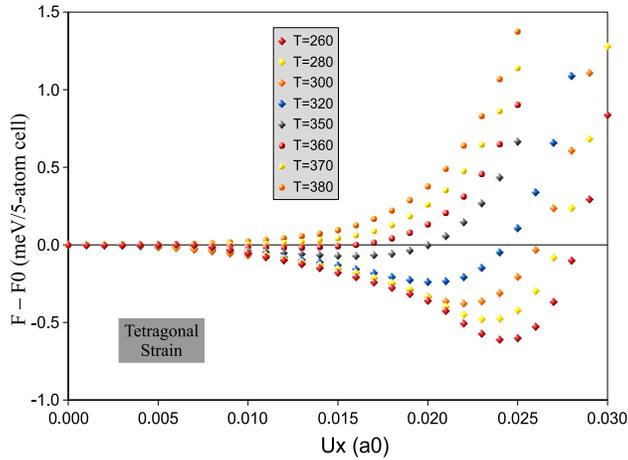}}}
    \par}
     \caption{{\small Landau free energy of tetragonal BTO (fixed volume) along [100] for 8 temperatures around the phase transition.}}
    \label{tetra1}
\end{figure}

We also calculate under this fixed strain tensor the free energy as a function of $u_x$ (with $u_y = u_z = 0.0$), for eight temperatures (from 260 to 380K) and plot it on Fig.~\ref{tetra1}.
The values of $u_x$ directly found by molecular dynamics correspond to the minima of the Landau free energy curve, as expected (the values minimizing the free energy are reported on Fig.~\ref{tetra2}, in orange circles). Interestingly, the transition is also second order or weakly first-order: within the numerical accuracy of our calculations, the evolution of the Landau free energy curves as a function of temperature follow the well-known trends of second order transitions, with a continuous evolution from a double-well to a single-well (with minimum at zero) curve.

In the range of temperature considered (260-400K), we do not find any orthorhombic phase. Such a phase naturally appears when the simulation is performed under fixed pressure and not under fixed volume\cite{zhong95,md}.

\begin{figure}[htbp]
    {\par\centering
    {\scalebox{0.30}{\includegraphics{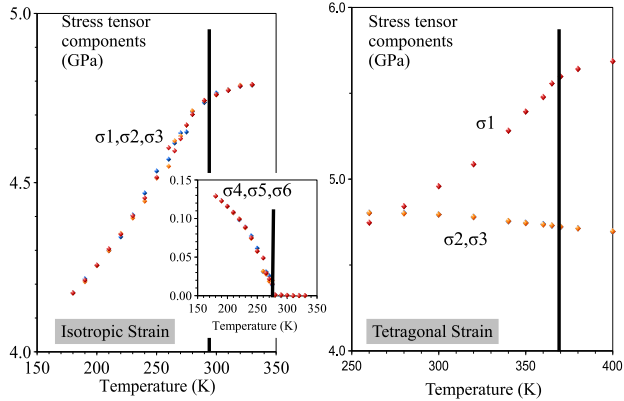}}}
    \par}
     \caption{{\small Temperature evolution of the stress tensor in the two cases studied under fixed volume (cubic and tetragonal). The strain is frozen in both cases to values obtained from direct MD simulations at a negative hydrostatic pressure -4.8 GPa ({\it i.e.} positive stress tensor), which explains why the stress tensor diagonal components have very high values around 4.8 GPa.}}
    \label{stress}
\end{figure}

The second-order (or weakly first-order) character of the two phase transitions is illustrated by the temperature evolution of the stress tensor components (Fig.~\ref{stress}), that do not exhibit discontinuities at the phase transition.

\subsection{Comparison with phenomenological potentials}

Since the work of Devonshire, the thermodynamics of barium titanate has been the subject of many phenomenological studies, and several Landau potentials have been proposed as expansions in power series of $\vec P$ up to 6$^{th}$ or 8$^{th}$ order for this material. The Landau potential is more often given as a Gibbs free energy and for a stress-free crystal ($\sigma$=0).

Converting to an Helmholtz free energy\cite{hlinka2006,nambu1994}, $F - F_0 (=\Delta F)$ usually writes as the sum of three terms, in which it appears as a function of temperature, strain tensor $\eta$ and polarization:

\begin{equation}
\Delta F(\eta,T;\vec P) = F_L(T;\vec P) + F_{el}(\eta,T) + F_c(\eta,T;\vec P),
\end{equation}

in which $\eta$ denotes the strain tensor (given in Voigt notations) and 

\begin{eqnarray*}
&& F_L(T;\vec P) = \alpha_{1}(P_x^2 + P_y^2 + P_z^2) + \alpha_{11}^{\eta} (P_x^4 + P_y^4 + P_z^4)  \\
&& + \alpha_{12}^{\eta}(P_x^2 P_y^2 + P_x^2 P_z^2 + P_y^2 P_z^2) + \alpha_{111} (P_x^6 + P_y^6 + P_z^6) \\
&& + \alpha_{112}(P_x^2(P_y^4+P_z^4) + P_y^2(P_x^4+P_z^4) + P_z^2(P_x^4+P_y^4)) \\
&& +\alpha_{123}P_x^2 P_y^2 P_z^2 + \alpha_{1111}(P_x^8 + P_y^8 + P_z^8)  \\
&& +\alpha_{1112}(P_x^6(P_y^2+P_z^2) + P_y^6(P_x^2+P_z^2) + P_z^6(P_x^2+P_y^2))  \\
&& + \alpha_{1122}(P_x^4 P_y^4 + P_x^4 P_z^4 + P_y^4 P_z^4)  \\
&& +\alpha_{1123}(P_x^4 P_y^2 P_z^2 + P_y^4 P_x^2 P_z^2 + P_z^4 P_x^2 P_y^2)
\end{eqnarray*}


\begin{eqnarray*}
&& F_{el}(\eta,T) = \frac{C_{11}}{2}(\eta_1^2 + \eta_2^2 + \eta_3^2) \\
&& + C_{12}(\eta_1 \eta_2 + \eta_1 \eta_3 + \eta_2 \eta_3) + \frac{C_{44}}{2} (\eta_4^2 + \eta_5^2 + \eta_6^2), \\
\end{eqnarray*}

where the $C_{ij}$ are the elastic constants,

\begin{eqnarray*}
&& F_{c}(\eta,T;\vec P) = -q_{11}(\eta_1 P_x^2 + \eta_2 P_y^2+ \eta_3 P_z^2)  \\
&& -q_{12} (\eta_1 (P_y^2 + P_z^2) + \eta_2 (P_x^2 + P_z^2) + \eta_3 (P_x^2 + P_y^2))  \\
&& -q_{44} (\eta_6 P_x P_y + \eta_5 P_x P_z + \eta_4 P_y P_z),  \\
\end{eqnarray*}

where the $q_{ij}$ are the electrostrictive coefficients.

The coefficients entering the expression of $F_L$ are the same as those entering the free energy of the stress-free crystal, excepting the 4$^{th}$-order coefficients, that must be renormalized\cite{hlinka2006,nambu1994}. These renormalized 4$^{th}$-order coefficients $\alpha_{11}^{\eta}$ and $\alpha_{12}^{\eta}$ are calculated from $\alpha_{11}$ and $\alpha_{12}$ (from the stress-free crystal) and from the elastic and electrostrictive coefficients\cite{hlinka2006}. In the following, we will use the values of Ref.~\onlinecite{hlinka2006} to calculate these renormalized coefficients.

This function $F-F_0$ should be directly comparable to the free energy we have calculated by thermodynamic integration. We will restrain to the comparison with two recent studies that have proposed 8$^{th}$-order expansions of the Landau free energy: Li {\it et al}\cite{li2005} and Wang {\it et al}\cite{wang2006,wang2007}. However, the reader must be cautionned that the comparison is not always relevant for the following reasons:

(i) the strains have not the same reference in the two models: in the effective hamiltonian the strain is defined with respect to the cubic $Pm\bar{3}m$ LDA ground state (lattice constant 7.46 a.u.), which causes problems in the comparison of the quadratic coefficients of $F$. 

(ii) the effective hamiltonian has a Curie temperature $\approx$ 120 K lower than the true one. Thus the quadratic coefficient, that is expected to evolve with T as $a(T-T_0)$, will be naturally strongly shifted with respect to the phenomenological potentials, that reproduce much better the transition temperatures.

(iii) finally, some free energy curves can give very different coefficients according to the degree of the polynomial function with which they are fitted.

For these reasons, we have prefered to plot the quadratic coefficient as obtained from our calculations without correction. From a general point of view, the coefficients fitted on our data have orders of magnitude in good agreement with the phenomenological potentials, although their temperature evolution appears as more complex, as already pointed out by I$\tilde{n}$iguez {\it et al}\cite{iniguez2001}.

\subsubsection{Isotropic strain case}
In the first case studied (isotropic strain), we have $P_x = P_y = P_z$ and $\eta_1 = \eta_2 = \eta_3$. Thus the free energy writes:

\begin{eqnarray*}
&& \Delta F(\eta,T;P_x) = 3(\alpha_{1}- (q_{11} + 2q_{12})\eta_1) P_x^2  \\
&& + 3(\alpha_{11}^{\eta} + \alpha_{12}^{\eta}) P_x^4 + 3(\alpha_{111}+2\alpha_{112}+\alpha_{123}/3) P_x^6 \\ 
&& + 3(\alpha_{1111} + 2\alpha_{1112} + \alpha_{1122} + \alpha_{1123}) P_x^8 + F_{el} \\
\end{eqnarray*}

The curves of Fig.~\ref{cubique1} have been fitted on 8$^{th}$ order polynomial functions (without odd term).

\begin{figure}[htbp]
    {\par\centering
    {\scalebox{0.37}{\includegraphics{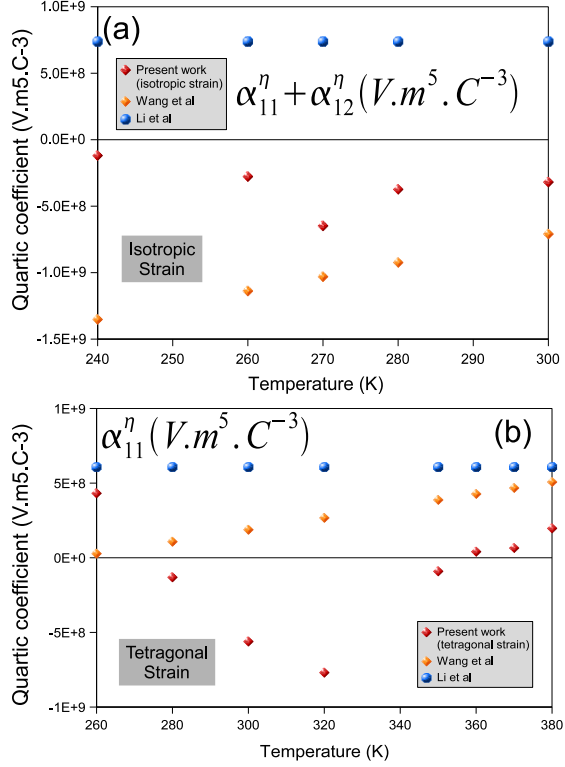}}}
    \par}
     \caption{{\small Temperature evolution of $\alpha_{11}^{\eta} + \alpha_{12}^{\eta}$ (a) and $\alpha_{11}^{\eta}$ (b) according to Wang {\it et al} (orange diamonds) and Li {\it et al} (blue circles). Red diamonds: coefficient of $P_x^4$, calculated from the isotropic strain case and divided by 3 (a), and calculated from the tetragonal strain case (b).}}
    \label{landau_quartic}
\end{figure}

The coefficients of the high-order terms give access to complex combinations of the $\alpha$ coefficients ($\alpha_{1111} + 2 \alpha_{1112} + \alpha_{1122} + \alpha_{1123}$ and $\alpha_{111} + 2 \alpha_{112} + \alpha_{123}/3$). The coefficient of $P_x^4$ is plotted on Fig.~\ref{landau_quartic}-a together with $\alpha_{11}^{\eta} + \alpha_{12}^{\eta}$ calculated from the potentials of Wang {\it et al} and {\it Li et al}. It falls just between the two and has the same sign as that of Wang.

%
%
%

\begin{figure}[htbp]
    {\par\centering
    {\scalebox{0.37}{\includegraphics{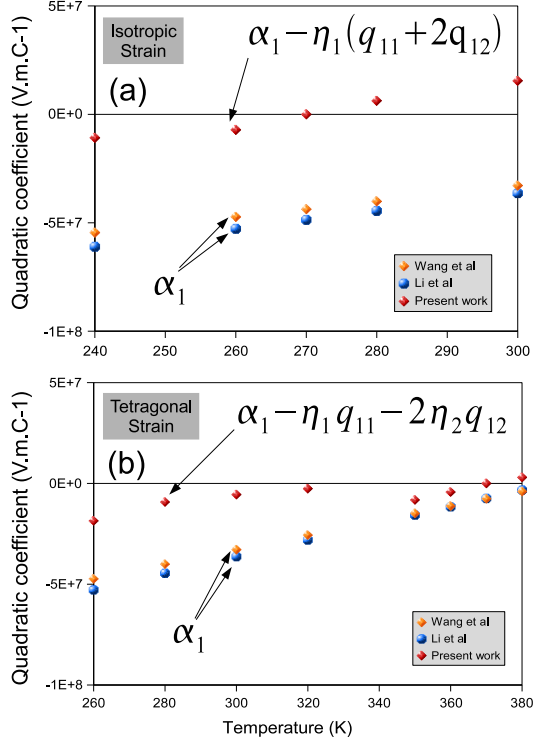}}}
    \par}
     \caption{{\small Temperature evolution of $\alpha_{1}$ (a and b) according to Wang {\it et al} (orange diamonds) and Li {\it et al} (blue circles). Red diamonds: coefficient of $P_x^2$, calculated from the isotropic strain case (a), and calculated from the tetragonal strain case (b).}}
    \label{landau_quadratic}
\end{figure}

The coefficient of $P_x^2$ is plotted on Fig.~\ref{landau_quadratic}-a as well as $\alpha_{1}$ calculated from the potentials of Wang {\it et al} and {\it Li et al}. As already said, they are not directly comparable since under fixed strain, $\alpha_{1}$ should be renormalized by a constant term $\alpha_{1} - \eta_1 (q_{11} + 2q_{12})$. Anyway the slopes are close. 
The term found from our calculation approximately varies with T as $a(T-T_0)$ with $T_0 \approx$ 270K, which roughly corresponds to the transition temperature in the isotropic strain case.

\subsubsection{Tetragonal case}
In the tetragonal case, we have $P_y = P_z = 0$ and $\eta_2 = \eta_3$. Thus, the Landau free energy writes:

\begin{eqnarray*}
&& \Delta F(\eta,T;P_x) = \alpha_{1} P_x^2 + \alpha_{11}^{\eta} P_x^4 + \alpha_{111} P_x^6 + \alpha_{1111} P_x^8 \\
&&  - q_{11} \eta_1 P_x^2 -2q_{12} \eta_2 P_x^2  + F_{el} \\
\end{eqnarray*}

Then, as in the previous section, we have fitted the curves of Fig.~\ref{tetra1} obtained by the thermodynamic integration by polynomial functions up to 8$^{th}$ order. For each temperature, we get a set of 4 coefficients that we now compare to the ones of Wang {\it el al} and Li {\it et al}.

\begin{figure}[htbp]
    {\par\centering
    {\scalebox{0.37}{\includegraphics{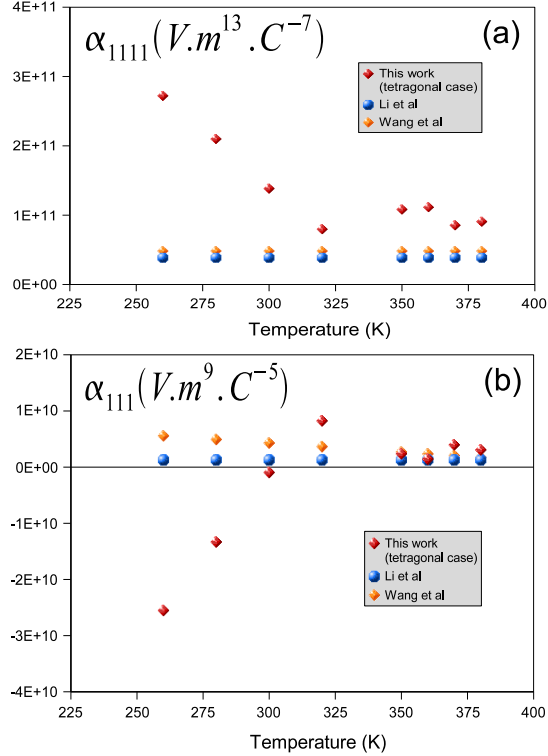}}}
    \par}
     \caption{{\small Temperature evolution of the $\alpha_{1111}$ and $\alpha_{111}$ coefficients according to Wang {\it et al} (orange diamonds) and Li {\it et al} (blue circles). Red diamonds: coefficients of the $P_x^8$ (a) and $P_x^6$ (b) terms as obtained from our calculations in the tetragonal strain case.}}
    \label{landau_tetra1}
\end{figure}

The highest order terms should directly correspond to the $\alpha_{111}$ and $\alpha_{1111}$ coefficients. The values provided by our data correspond very well to the phenomenological ones above $\approx$ 325 K (Fig.~\ref{landau_tetra1}). Below this temperature, these coefficients strongly vary.

The variation of the $P_x^4$ coefficient, that should compare directly to $\alpha_{11}^{\eta}$, reveals more complex (Fig.~\ref{landau_quartic}-b), as well as that of the quadratic one (Fig.~\ref{landau_quadratic}-b).

\section{Conclusion}
In this work, we have proposed and validated a method to calculate numerically the Landau free energy of a ferroelectric material, without any assumption on its analytical form and using the effective hamiltonian of Zhong {\it et al}\cite{zhong95} directly fitted on first-principles calculations. It is based on the technique of the thermodynamic integration coupled with Molecular Dynamics performed at fixed temperature (Nos\'e-Hoover) and fixed polarization along paths in the 3-D space (P$_x$, P$_y$, P$_z$). Although possible also with Monte Carlo simulations, Molecular Dynamics is well suited to this kind of problem since thermodynamic integration requires to calculate the forces, directly available within MD.

The method has been applied to the case of BTO under fixed volume and showed that minima of the calculated Landau free energy correspond to the results of direct MD simulations. The results have been compared with some available phenomenological potentials.

This method connects directly the phenomenological theories to the atomic-scale simulations. The next step is to extend the simulations to the isothermal-isobaric ensemble to obtain the Gibbs free energy as a function of polarization. This will be achieved by coupling Nos\'e-Hoover, Parinello-Rahman and also fixed polarization molecular dynamics. Other extensions of this approach can be imagined, such as the computation of entropy profiles as a function of $\vec P$, the application to low-dimensional systems and to complex perovskite (with antiferrodistortions) and multiferroic materials, where the free energy could be computed as a function of polarization and tilt angle or magnetization\cite{kornev2007}.

The method also offers a wide freedom to use any internal parameter of the system as variable to expand the free energy\cite{darve2001}.

%

\section{Appendix A: the Landau free energy as a potential of mean force}

In this section we give proofs for Eq.~\ref{int_thermo}, very similar to demonstrations that can be found in the chemical scientific litterature on thermodynamic integration\cite{kirkwood,mezei,ti1,darve2001}.
Let us considere two macroscopic states of the system (1) and (2), each defined by (N,V,T), but having different polarizations $\vec P$ = $Z^{*}/\Omega \sum_i \vec u_i = N Z^{*}/\Omega \vec u$. It is more convenient to work with the average local mode $\vec u$, related to $\vec P$ only by a constant factor.

We introduce as usual an incomplete free energy, for a fixed value of $\vec u$, defined by
\begin{equation}
\tilde{F}(N,V,T,\vec u) = -k_B T ln \tilde{Z}(N,V,T,\vec u),
\end{equation}
$\tilde{Z}(N,V,T,\vec u)$ being an incomplete partition function defined by a summation over the microscopic states having a polarization equal to $\vec P = N Z^{*}/\Omega \vec u$.

The difference of free energy between state (1) and state (2) is
\begin{equation}
\tilde{F}(N,V,T,\vec u_2) - \tilde{F}(N,V,T,\vec u_1) = \oint \left\{ \frac{\partial \tilde{F}}{\partial \vec u} \right\}_{N,V,T}.d \vec u,
\end{equation}
in which the derivative with respect to a vector denotes the gradient operator $\vec \nabla_{\vec u} = (\partial/\partial u_x, \partial/\partial u_y, \partial/\partial u_z)$. $\oint$ is the integral over a continuous path joining (1) to (2) in the 3-D space ($u_x$, $u_y$, $u_z$).

We have thus to calculate the term 
\begin{equation}
\frac{\partial \tilde{F}}{\partial \vec u} = - \frac{k_B T}{\tilde{Z}} \frac{\partial \tilde{Z}}{\partial \vec u}
\end{equation}

We start by calculating the incomplete partition function $\tilde{Z}(N,V,T,\vec u)$
\begin{equation}
\tilde{Z}(N,V,T,\vec u) = \sum_{r/ \vec u} e^{- \beta E_r}
\end{equation}

$(r)$ denoting the microscopic states and the summation being over the $(r)$ with polarization $\vec P = N Z^{*}/\Omega \vec u$ and $\beta = 1/k_B T$.
We use the classical approximation to calculate $\tilde{Z}$ and omit in the following N,V,T in the notations, all the partial derivatives being performed at constant $N,V,T$.
In this framework, the partition function is an integral over the phase space, the summation being restricted to the microscopic states satisfying the constraint $\sum_i \vec u_i = N \vec u$. It writes\cite{otter,darve2001}:

\begin{eqnarray*}
&&  \tilde{Z}(\vec u) = \frac{1}{h^{3N}} \int d\vec p_1 ...d\vec p_N \int d\vec u_1 ...d\vec u_{N} \\
&& \times \delta(\sum_i \vec u_i - N \vec u) e^{-\beta E(\vec p_1 ...\vec p_N,\vec u_1 ...\vec u_{N})},
\end{eqnarray*}

In this expression, the energy of the microstates is:
\begin{eqnarray*}
&& \mathrm E(\vec p_1 ...\vec p_N,\vec u_1 ...\vec u_{N}) = \sum_{i=1}^N \frac{\vec p_i^2}{2 m_{lm}} + 
H^{eff}(\vec u_1 ...\vec u_{N})
\end{eqnarray*}

The $\vec p_i$ are the conjuguate momenta of the local modes $\vec u_i$. In the expression of $\tilde{Z}$, $N!$ is not introduced since the local modes, which correspond to well identified unit cells, are discernable. Moreover we do not write the displacement modes $\vec v_i$ for simplicity.

Integrating over the $\vec p_i$ yields:

\begin{eqnarray*}
\label{integral}
&&  \tilde{Z}(\vec u) =  \left\{ \frac{2 \pi m k_B T}{h^2} \right\}^{3N/2} \int d\vec u_1 ...d\vec u_{N} \\
&& \times \delta(\sum_i \vec u_i - N \vec u) e^{-\beta H^{eff}(\vec u_1 ...\vec u_{N})}
\end{eqnarray*}

This integral can be calculated in several manners. One can remove the Dirac function by replacing one of the $\vec u_i$ by $N \vec u - \sum_{j \neq i} \vec u_j$. For instance,

\begin{eqnarray*}
&&  \tilde{Z}(\vec u) =  \left\{ \frac{2 \pi m k_B T}{h^2} \right\}^{3N/2} \int d\vec u_1 ...d\vec u_{N-1} \\
&& \times  e^{-\beta H^{eff}(\vec u_1, ...\vec u_{N-1}, N \vec u - \sum_{i=1}^{N-1} \vec u_i)}
\end{eqnarray*}
The microscopic states satisfying the constraint thus appear as functions of $\vec u$ and indexed by $\vec u_1, \vec u_2, ... \vec u_{N-1},\vec p_1 ... \vec p_N$. This is more obvious by rewriting:
$H^{eff}_{\vec u_1 ...\vec u_{N-1}}(\vec u) = H^{eff}(\vec u_1 ...\vec u_{N-1},N \vec u - \sum_{i=1}^{N-1} \vec u_i)$,

\begin{eqnarray*}
&&  \tilde{Z}(\vec u) =  \left\{ \frac{2 \pi m k_B T}{h^2} \right\}^{3N/2} \int d\vec u_1 ...d\vec u_{N-1} \\
&& \times  e^{-\beta H^{eff}_{\vec u_1 ...\vec u_{N-1}}(\vec u)}
\end{eqnarray*}

The derivative with respect to $\vec u$ is:
\begin{eqnarray*}
&&  \frac{\partial \tilde{F}}{\partial \vec u} =  - \frac{k_B T}{\tilde{Z}} \frac{\partial \tilde{Z}}{\partial \vec u} = \\
&&  \int d\vec u_1...d\vec u_{N-1} \frac{\partial H^{eff}_{\vec u_1...\vec u_{N-1}}(\vec u)}{\partial \vec u} \\
&& \times \frac{ e^{-\beta H^{eff}_{\vec u_1...\vec u_{N-1}}(\vec u)}}
{\int d\vec u_1...d\vec u_{N-1} e^{-\beta H^{eff}_{\vec u_1...\vec u_{N-1}}(\vec u)}},
\end{eqnarray*}

which can be rewritten:

\begin{eqnarray*}
&& \frac{\partial \tilde{F}}{\partial \vec u} = \frac{1}{h^{3N}} \int d \vec p_1 ... d \vec p_N
\int d\vec u_1...d\vec u_{N}  \delta(\sum_i \vec u_i - N \vec u) \\
&& \times \frac{\partial H^{eff}_{\vec u_1...\vec u_{N-1}}(\vec u)}{\partial \vec u} \frac{ e^{-\beta E_{\vec p_1 ... \vec p_N,\vec u_1 ... \vec u_{N}}(\vec u)}}
{\tilde{Z}}
\end{eqnarray*}

One recognizes the thermal average (under fixed polarization) in the canonical ensemble of the microscopic quantity $\frac{\partial H^{eff}_{\vec u_1...\vec u_{N-1}}(\vec u)}{\partial \vec u}$.

This quantity is in fact: 
\begin{equation}
\frac{\partial H^{eff}_{\vec u_1...\vec u_{N-1}}(\vec u)}{ \partial \vec u} = N \frac{\partial H^{eff}}{\partial \vec u_N}
= -N \vec f_N,
\end{equation}
in which $\vec f_N$ is the force acting on the N$^{th}$ local mode.

And finally,

\begin{equation}
\frac{\partial \tilde{F}}{\partial \vec u} = -N \left\langle \vec f_N \right\rangle
\end{equation}
It is clear that the N$^{th}$ local mode does not have a particular role. It appears here because of the choice made to calculate the initial integral. If we had chosen to remove $\vec u_1$ instead of $\vec u_N$, we would have found
\begin{equation}
\frac{\partial \tilde{F}}{\partial \vec u} = -N \left\langle \vec f_1 \right\rangle
\end{equation}

In fact, the system is completely homogeneous and the thermal average of the forces does not depend on the site on which it is calculated:
\begin{equation}
\left\langle \vec f_1 \right\rangle = \left\langle \vec f_2 \right\rangle = ...
=\left\langle \vec f_N \right\rangle
\end{equation}

To improve the statistics of the thermal average, it is more interesting to calculate the thermal averages of all the forces and to sum them:
\begin{equation}
\left\{ \frac{\partial \tilde{F}}{\partial \vec u} \right\}_{N,V,T} = - \sum_{i=1}^N \left\langle \vec f_i \right\rangle
\end{equation}
The thermal average in this expression is performed at fixed polarization.
The quantity that must be integrated is thus the mean force acting on the "reaction coordinate", {\it i.e.} the polarization (this is the total force that has to be added in the equations of motion to maintain the polarization constant). The free energy is thus the potential of the mean force acting on the polarization.

We note that in some cases, the use of more complex reaction coordinates can give rise to more complex expressions for the derivative of the free energy\cite{otter,darve2001}. This is not the case here.

Very similar arguments in the isothermal-isobaric ensemble lead to
\begin{equation}
\label{ggg}
\left\{ \frac{\partial \tilde{G}}{\partial \vec u} \right\}_{N,P,T} = - \sum_{i=1}^N \left\langle \vec f_i \right\rangle,
\end{equation}

which is formally the same expression, except that the thermal average must be performed under fixed temperature, {\it pressure} and $\vec u$. This simple extension is related to the fact that local modes and homogeneous strain are independent variables of the Effective Hamiltonian.

In term of polarization (instead of average local mode), it writes:
\begin{equation}
\label{ggg2}
\left\{ \frac{\partial \tilde{G}}{\partial \vec P} \right\}_{N,P,T} = - \frac{\Omega}{NZ^{*}} \sum_{i=1}^N \left\langle \vec f_i \right\rangle,
\end{equation}

The calculation of $\frac{\partial \tilde{G}}{\partial \vec u}$ could be performed by using the Parinello-Rahman\cite{pr} scheme combined to Nos\'e-Hoover and fixed polarization.

\section{Appendix B: Molecular Dynamics using Effective Hamiltonians}
The effective hamiltonian is solved in our approach by Molecular Dynamics\cite{md}. We describe here the main features of this approach. A specific mass is affected to the local modes and displacement modes, and these degrees of freedom evolve in time according to the Newton equations of motion. At each time, the forces acting on these local modes derive from the effective hamiltonian:

\begin{equation}
m_{lm}\frac{d^2 \vec u_i}{dt^2} = - \frac{\partial H^{eff}}{\partial \vec u_i} = \vec f_i^{lm}
\end{equation}

\begin{equation}
m_{dsp}\frac{d^2 \vec v_i}{dt^2} = - \frac{\partial H^{eff}}{\partial \vec v_i} = \vec f_i^{dsp}
\end{equation}

where $lm$ and $dsp$ stand respectively for local mode and displacement mode. These equations are integrated numerically within the well-known Verlet algorithm\cite{verlet}: the local mode at time $t+h$ ($h$ is the time step) is computed from the local modes at $t$ and $t-h$, using the forces calculated at $t$:

\begin{equation}
\vec u_i(t+h) = 2 \vec u_i(t) - \vec u_i(t-h) + \frac{h^2}{m_{lm}} \vec f_i^{lm}(t)
\end{equation}

and idem for the displacement modes.

To fix the temperature, we have implemented the Nos\'e-Hoover method\cite{nose84,nose86,hoover85}. This method consists in adding to the system a fictitious degree of freedom $s$ representing the thermostat, and that couples to the equations of motion, according to:

\begin{equation}
\nonumber
m_{lm}\frac{d^2 \vec u_i}{dt^2} =  \vec f_i^{lm} - m_{lm} \zeta (t) \frac{d \vec u_i}{dt}
\end{equation}

\begin{equation}
\nonumber
m_{dsp}\frac{d^2 \vec v_i}{dt^2} =  \vec f_i^{dsp} - m_{dsp} \zeta (t) \frac{d \vec v_i}{dt}
\end{equation}

\begin{equation}
\nonumber
\frac{d \zeta}{dt} =  - \frac{k_B g T(t)}{Q} (\frac{T_0}{T(t)} -1)
\end{equation}
where T(t) and T$_0$ are respectively the instantaneous and targetted temperature, $g$ is the number of degrees of freedom in the system, and $\zeta (t)= \frac{d ln s}{dt} (t)$. $Q$ is the so-called Nos\'e mass, that must be chosen correctly so that the exchange of energy between the system and the thermostat are neither too slow nor too fast. The velocities at time step $t$ are calculated from the Ferrario algorithm\cite{ferr85}:

\begin{equation}
\label{vel-ferr85}
\frac{d \vec u_i}{dt}(t) = \frac{3 \vec u_i(t) -4 \vec u_i(t-h) + \vec u_i(t-2h)}{2 h}
\end{equation}

\section{Appendix C: Relationship to thermodynamics}
The basic laws of thermodynamics allow to write:

\begin{equation}
dF = \delta W' - PdV -SdT
\end{equation}

where $\delta W'$ is the work of external forces other than pressure forces, and $-PdV$ is the work of pressure forces.
A system evolving reversibly under constant N,V and T obeys: $dF = \delta W'$.

In our case, the variation of the mean local mode from $\vec u$ to $\vec u + d\vec u$ (which has the dimension of a distance) is related to the work of the force $- \sum_i \left\langle \vec f_i^{lm} \right\rangle$, {\it i.e.}:

\begin{equation}
dF = \delta W' = - \sum_i \left\langle \vec f_i^{lm} \right\rangle . d\vec u
\end{equation}

The Landau free energy variation is thus the work of (minus) the mean force acting on the polarization (this is the external force added to maintain the constraint in the molecular dynamics).

For the Gibbs free energy:
\begin{equation}
dG = \delta W' + VdP -SdT
\end{equation}
and a system evolving reversibly under constant N,P and T obeys: $dG = \delta W'$. Thus in such conditions

\begin{equation}
dG = \delta W' = - \sum_i \left\langle \vec f_i^{lm} \right\rangle . d\vec u
\end{equation}

\end{document}